\documentclass{article}
\usepackage{spconf}
\usepackage{amsmath,graphicx}
\usepackage{tabularx}
\usepackage{multirow}
\usepackage{amssymb}
\usepackage{verbatim}
\usepackage{color}
\usepackage{caption}
\usepackage{cite}
\usepackage{url}
\usepackage{float}
\usepackage{mathtools}
\usepackage[bottom]{footmisc}

\title{Learned Convolutional Sparse Coding}

\name{Hillel Sreter\sthanks{This research is supported by Wipro ltd. One of the GPUs used for this research was donated by the NVIDIA Corporation.} \qquad Raja Giryes}

\address{ School of Electrical Engineering, Tel Aviv University,
Tel Aviv, Israel,
 \{hillelsr@mail., raja@\}tau.ac.il }

\begin{document}

    \maketitle
    \begin{abstract}
        We propose a convolutional recurrent sparse auto-encoder model. The model consists of a sparse encoder, which is a convolutional extension of the learned ISTA (LISTA) method, and a linear convolutional
        decoder.
        Our strategy offers a simple method for learning a task-driven sparse convolutional dictionary (CD),
        and producing an approximate convolutional sparse code (CSC) over the learned dictionary.
        We trained the model to minimize reconstruction loss via gradient decent
        with back-propagation and have achieved competitve results to KSVD image denoising and to leading CSC methods in image inpainting requiring only a small fraction of their run-time.
    \end{abstract}


    \section{Introduction}
    \label{sec:intro}
        Sparse coding (SC) is a powerful and popular tool used in a variety of applications
        from classification and feature extraction,
        to signal processing tasks such as enhancement, super-resolution, etc.
        A classical approach to  use SC with a signal is to
        split it into segments (or patches), and solve for each
        \begin{equation}\label{eq:1}
		\min || \mathbf{z} ||_0\; s.t.\; \mathbf{x}=\mathbf{D}\mathbf{z},
        \end{equation}

        \noindent where $\mathbf{z} \in \mathbb{R}^m$ is the sparse representation of the (column stacked) patch $\mathbf{x}\in \mathbb{R}^n$ in the dictionary $\mathbf{D} \in \mathbb{R}^{n\times m}$.
         There are two painful points in this approach: (i)
        performing SC over all the patches tends to be a slow
        process; and (ii) learning a dictionary
        over each segment independently
        loses spatial information outside it such as shift-invariance in images.\par One prominent approach  for addressing the first point
        is using approximate sparse coding models such as the learned iterative shrinkage and thresholding algorithm (LISTA)
         \cite{gregor2010learning}.
        LISTA is a recurrent neural network architecture designed to mimic ISTA \cite{daubechies2004iterative}, which is an iterative algorithm for approximating the solution of (\ref{eq:1}).
        As to the second point, one may impose an additional prior on the learned dictionary such as being convolutional, i.e., a concatenation of Toeplitz matrices.
        In this case each element (known also as atom) of the dictionary is learned based on the whole signal.
        Moreover, the resulting dictionary is shift-invariant due to it being convolutional.\par
        In this paper, we introduce a learning method for task driven  CD learning. We design a convolutional neural network that both learns the SC of a family of signals and their CSCs. We demonstrate the efficiency of our new approach in the tasks of image denoising and inpainting.
        \section{The Sparse Coding Problem}
        \label{sec:scproblem}
        \subsection{Sparse coding and ISTA}
        Solving (\ref{eq:1}) directly is a combinatorial problem, thus, its complexity grows
        exponentially with $m$. To resolve this deficiency, many approximation strategies have been developed for solving (\ref{eq:1}). A popular one relaxes the $\ell_0$ pseudo-norm using the $\ell_1$ norm yielding (the unconstrained form): \\
        \begin{equation}\label{eq:2}
        \arg\min_{\mathbf{D}, \mathbf{z}} \frac{1}{2}||\mathbf{x} - \mathbf{D}\mathbf{z}||_2^2 + \lambda||\mathbf{z}||_1.
        \end{equation}
A popular technique to minimize (\ref{eq:2}) is the ISTA \cite{daubechies2004iterative} iteration:
        \begin{equation}\label{eq:3}
        \mathbf{z_{k+1}} = S_{\lambda/L}(\mathbf{z_k} + \frac{1}{L}\mathbf{D}^T(\mathbf{D}\mathbf{x}-\mathbf{z})),
        \end{equation}
        where $L \leq \sigma_{max}(\mathbf{D}^T\mathbf{D})$, $\sigma_{max}(\mathbf{A})$
        is the largest eigenvalue of $\mathbf{A}$ and  $S_\theta(x)$ is the soft thresholding operator:
        \begin{equation}\label{eq:4}
          S_\theta(x)=sign(x)max(|x|-\theta, 0).
        \end{equation}
        ISTA iterations are stopped when a defined convergence criterion is satisfied.
        \subsection{Learned ISTA (LISTA)}
		In approximate SC, one may build a non-linear differentiable encoder that can be trained to produce
		quick approximate SC of a given signal. In \cite{gregor2010learning}, an ISTA like recurrent network denoted by LISTA
		is introduced. By rewriting (\ref{eq:3}) as
		\begin{equation}\label{eq:5}
		\mathbf{z_{k+1}} = S_{\lambda/L}((\mathbf{I}-\mathbf{D}^T\mathbf{D})\mathbf{z_k}+\mathbf{D}^T\mathbf{x}),
		\end{equation}

		\noindent LISTA can be derived by substituting in (\ref{eq:5}) $\mathbf{D}^T$ for $\mathbf{W_e} \in \mathbb{R}^{m\times n}$, $I - \frac{1}{L}\mathbf{D}^T\mathbf{D}$ for $\mathbf{S} \in \mathbb{R}^{m \times m}$ and $\frac{\lambda}{L}$ for $\theta \in \mathbb{R}_{+}^m$ (using a separate threshold value for each entry instead of a single threshold for all values as in \cite{sprechmann2015learning}).
        Thus, LISTA iteration is defined as:
		\begin{equation}\label{eq:6}
		\begin{split}
		& \mathbf{z}_0 = 0, \;\; k=0..K-1\\
		& \mathbf{z}_{k+1}= S_\theta(\mathbf{S}\mathbf{z}_k + \mathbf{W_e}\mathbf{x}) \\
		&\mathbf{z}_{LISTA} = \mathbf{z}_K,
		\end{split}
		\end{equation}
		where $K$ is the number of LISTA iterations and the parameters $\mathbf{W_e},\,\mathbf{S}$ and $\theta$ are learned by minimizing:
		\begin{equation}\label{eq:7}
		\mathcal{L} = \frac{1}{2}||\mathbf{z}^*- \mathbf{z}_{LISTA}||_2^2,
		\end{equation}
		where $\mathbf{z}^*$ is either the optimal SC of the input signal (if attainable) or the ISTA final solution after its convergence.
		\par There have been lots of work on LISTA like architectures. In \cite{sprechmann2015learning}, a LISTA auto-encoder is introduced. Rather than training to approximate the optimal SC, LISTA is trained directly to minimize the optimization objective (\ref{eq:2}) instead of (\ref{eq:7}).
A discriminative non-negative version of LISTA is introduced in \cite{rolfe2013discriminative}.
		It contains a decoder that reconstructs the approximate
		SC back to the input signal	and a classification unit that receives the approximate SC as an input.
		Thus, the model described in \cite{rolfe2013discriminative}  is named discriminative recurrent sparse auto-encoder and the quality of the produced
		approximate SC is quantified by the success of the decoder and classifier.
	    The work in \cite{wang2015deep} used a cascaded sparse coding network with LISTA building blocks to fully exploit the sparsity of an image for the super resolution problem.
	    In \cite{giryes2016tradeoffs} and \cite{moreau2016adaptive}, a theoretical explanation is provided for the reason LISTA like models are able to accelerate SC iterative solvers.
    \section{Convolutional Sparse Coding}
    The CSC model \cite{heide2015fast},\cite{bristow2013fast},\cite{papyan2017convolutional}, \cite{bristow2014optimization}, \cite{zeiler2010deconvolutional}  can be derived from the classical SC model by  substituting matrix multiplication with the convolution operator:
    \begin{equation}\label{eq:8}
        \mathbf{x} = \sum_{i=0}^{m-1}\mathbf{d}_i * \mathbf{z}_i,
    \end{equation}
    where $ \mathbf{x} \in \mathbb{R}^{n_1 \times n_2}$ is the input signal,
    $\mathbf{d}_i \in \mathbb{R}^{k \times k}$ a local convolution filter and $\mathbf{z}_i \in \mathbb{R}^{n_1 \times n_2}$ a sparse feature map of the convolutional atom $\mathbf{d}_i$.
    The $\ell_1$ minimization problem in (\ref{eq:2}) for CSC may be formulated as:
 \begin{equation}\label{eq:9}
		\begin{split}
            & \arg\min_{\mathbf{d},\mathbf{z}} \frac{1}{2}||\mathbf{x}-\sum_{i=0}^{m-1}\mathbf{d}_i * \mathbf{z}_i||_2^2 + \lambda \sum_{i=0}^{m-1}||\mathbf{z}_i||_1.
		\end{split}
\end{equation}
    It is important to note that unlike traditional SC, $\mathbf{x}$ is not split into patches (or segments) but rather the CSC is of the full input signal.
 The  CSC model is inherently spatially invariant, thus, a learned atom of a specific edge orientation can globally represent all edges of that orientation in the whole image.
 Unlike CSC, in classical SC multiple atoms tend to learn the same oriented edge for different offsets in space. \par Various methods have been proposed for solving (\ref{eq:9}). The strategies in  \cite{heide2015fast} and\cite{bristow2013fast} involve transformation to the frequency domain and optimization using Alternating Direction Method of Multipliers (ADMM). These methods tend to optimize over the whole train-set at once. Thus, the whole train-set must be held in memory while learning the CDC, which of course limits the train-set size. Moreover, inferring $\mathbf{z}_i$ from $\mathbf{x}$ is an iterative process that may require a large number of iterations, thus, less suitable for real-time tasks. Work on speeding up the ADMM method for CD learning is done in\cite{wamg2017online}. In \cite{choudhury2017consensus}, a consensus-based optimization framework has been proposed that makes CSC tractable on large-scale
datasets. In \cite{garcia2017convolutional}, a thorough performance comparison among different CD learning methods is done as well as proposing a new learning approach. More work on CD learning has been done in  \cite{papyan2017convolutional} and \cite{wohlberg2014efficient}. In \cite{wohlberg2016boundary}, the effect of solving in the frequency domain on boundary artifact is studied, offering different types of solutions.The work in \cite{Gu_2015_ICCV} shows the potential of CSC for image super-resolution.
    \section{Learned Convolutional Sparse Coding}
    In order to have computationally efficient CSC model, we extend the approximate SC model LISTA \cite{gregor2010learning} to the CSC model.
    We perform training in an end-to-end task driven fashion. Our proposed approach
    shows competitive performance to classical SC and CSC methods in different tasks but with order of magnitude fewer computations.
     Our model is trained via stochastic gradient-descent, thus, naturally it can learn the CD over very large datasets without any special adaptations. This of course helps in learning a CD that can better represent the space in which the input signals are sampled from.
        \subsection{Learning approximate CSC}
        Due to the linear properties of convolutions and the fact that CSC model can be thought of as a classical SC model, where the dictionary $\mathbf{D}_{conv}$ is a concatenation of Toepltiz matrices, the CSC  model can be viewed as a specific case of classical SC. Thus,the objective in (\ref{eq:9}) can be formatted to be like (\ref{eq:2}), by substituting the general dictionary $\mathbf{D}$ with $\mathbf{D}_{conv}$. Obviously, naively reformulating CSC model as matrix multiplication
        is very inefficient both memory-wise because $\mathbf{D}_{conv} \in \mathbb{R}^{(n_1n_2) \times (n_1n_2m)}$ and computation wise, as each element of $\mathbf{x}$ is computed with $n_1n_2m$  multiply and accumulate operations (MACs) versus the convolution formation, where
        only $s^2m$ MACs are needed (realistically assuming $s \ll \{n_1, n_2\}$). \\
        \indent Thus, instead of using standard LISTA directly on $\mathbf{D}_{conv}$, we reformulate ISTA to the convolutional case and then propose its LISTA version. ISTA iterations for CSC reads as:
        \begin{equation}\label{eq:10}
            \mathbf{z}_{k+1} = S_{\lambda/L}(\mathbf{z}_k + \frac{1}{L}\mathbf{d} \star (\mathbf{x}-\mathbf{d}*\mathbf{z}_k)),
        \end{equation}
        where  $\mathbf{d} \in \mathbb{R}^{s \times s \times m}$ is an array of $m$ $s \times s$ filters, $\mathbf{d}\star\mathbf{x}=[flip(\mathbf{d}_0)*\mathbf{x},...,flip(\mathbf{d}_{m-1})*\mathbf{x}]$ and $\mathbf{d}*\mathbf{z}=\sum_{i=0}^{m-1}\mathbf{d}_i*\mathbf{z}_i$. The operartion $flip(\mathbf{d}_i)$ reverses the order of entries in $\mathbf{d}_i$ in both dimensions.
        Modeling (\ref{eq:10}) in a similar way to (\ref{eq:6}) (with some differences) leads to the convolutional LISTA structure:
    \begin{equation}\label{eq:11}
        \mathbf{z}_{k+1}= S_\theta(\mathbf{z}_k + \mathbf{w_e} * (\mathbf{x} - \mathbf{w_d}*\mathbf{z}_k)),
	\end{equation}
	where $\mathbf{w_e} \in \mathbb{R}^{s\times s \times c \times m}$, $\mathbf{w_d}\in \mathbb{R}^{s\times s \times m \times  c }$ and $\theta \in \mathbb{R}_{+}^{m}$ are fully trainable and independent
	variables. Note that we have added here also the variable $c$ that takes into account having multiple channels in the original signal (e.g., color channels).
     \subsection{Learning the CD}
     When learning a CD, we expect to produce $\hat{\mathbf{x}}$ as close as possible to $\mathbf{x}$
     given the approximate CSC (ACSC) produced by the model described in (\ref{eq:11}).
      Thus, we learn the CD by adding a linear encoder that consists of a
       filter array $\mathbf{d}$ at the end of the convolutional LISTA iterations.
       This leads to the following network that integrates both the calculation of the CSC and the application of the dictionary:
       \begin{equation}\label{eq:convlista}
		\begin{split}
		&\mathbf{z}_0 = 0, \;\;, k=0:K-1\\
		&\mathbf{z}_{k+1}= S_\theta(\mathbf{z}_k + \mathbf{w_e} * (\mathbf{x} - \mathbf{w_d}*\mathbf{z}_k)) \\
		&\mathbf{z}_{ACSC} = \mathbf{z}_K\\
		&\mathbf{\hat{x}}=\mathbf{d}*\mathbf{z}_{ACSC}
		\end{split}
		\end{equation}
     \subsection{Task driven convolutional sparse coding}
     This formulation makes it possible to train a sparse induced auto-encoder, where the encoder learns
    to produce ACSC and the decoder learns the correct filters to reconstruct the signal from the generated ACSC. The whole model is trained via stochastic gradient decent aiming at minimizing:
    \begin{equation}\label{loss}
    \mathcal{L}=dist(\mathbf{x}, \hat{\mathbf{x}}),
    \end{equation}
    where $\mathbf{x}$ is the target signal and $\mathbf{\hat{x}}$ is the one calculated in (\ref{eq:convlista}). We tested different types of distance functions and found (\ref{ssimloss}) to yield best results.
     Unlike the sparse auto-encoder proposed in \cite{ng2011sparse}, where the encoder
     is a feed-froward network and sparsity is induced via adding sparsity promoting regularization to the loss function, our model is inherently biased to produce sparse encoding of its input due to the special design of its architecture. From a probabilistic point of view, as shown in \cite{Goodfellow-et-al-2016}, sparse auto-encoder can be thought of as a
     generative model that has latent variables with a sparsity prior. Thus, the joint distribution of the model,output and CSC is
     \begin{equation}\label{eq:12}
		p_{model}(\mathbf{x}, \mathbf{z})=p_{model}(\mathbf{z})p_{model}(\mathbf{x}|\mathbf{z}).
	\end{equation}
	The soft thresholding operation used in the network encourages $p_{model}(\mathbf{z})$ to be large when $\mathbf{z}$ is sparse.
	Thus, when training our model we found it sufficient to minimize a reconstruction term representing
	$-log(p_{model}(\mathbf{x}|\mathbf{z}))$ without the need to add a sparsity inducing term to the loss.
    \section{EXPERIMENTS}
    	\subsection{Learned CSC network parameters}
    	We used our model as specified in (\ref{eq:convlista}). We found  3 recurrent steps to be sufficient for our tasks. The filters used in the experiments have the following dimensions:
    	$\mathbf{w_e} \in \mathbb{R}^{7 \times 7 \times 1 \times 64}$,
    	$\mathbf{w_d} \in \mathbb{R}^{7 \times 7 \times 64 \times 1}$,
    	$\theta \in \mathbb{R}_{+}^{1 \times 64}$,
    	$\mathbf{d} \in \mathbb{R}^{7 \times 7 \times 64 \times 1}$.\\
    	\indent As initialization is important for convergence,
    	$\mathbf{w_d}$ and $\mathbf{d}$ are initialized with the same random values and and
        $\mathbf{w_e}$ is initialized with the transposed and flipped filters
        of $\frac{1}{10}\mathbf{w_d}$. The factor $\frac{1}{10}$ takes into consideration $\frac{1}{L}$ from (\ref{eq:3}).
        We initialized the threshold $\theta$ to $\frac{1}{10}$, thus implicitly initializing
         $\lambda$ to $1$.
         We tested different types of reconstruction loss including standard $\ell_1$ and $\ell_2$ losses.
          We found the loss function proposed in \cite{zhao2017loss} to retrieves the best image quality,\\
      \begin{equation}\label{ssimloss}
    \mathcal{L}(\mathbf{x},\mathbf{\hat{x}})=\alpha (1 - ms\_ssim(\mathbf{x}, \mathbf{\hat{x}})) + (1-\alpha)||\mathbf{x} - \mathbf{\hat{x}}||_1,
    \end{equation}
   where $ms\_ssim$ is the multiscale SSIM loss.
    We have trained the model using the Adam optimizer \cite{Kingma2014adam},with this reconstruction loss
    and $\alpha=0.8$.
    We use PASCAL VOC \cite{pascal-voc-2012} dataset due to its large amount of high quality images. All images are normalized by 255 before feeding them to the model.
        \subsection{Image denoiseing}
    To test our model for image denoising we have added random noise to a given original image $\mathbf{x}$ producing $\mathbf{y}=\mathbf{x}+\epsilon$,
    where $\sigma_n=20$ and $\epsilon \sim \mathcal{N}(0,\,\sigma_n^2)$.

\begin{figure*}
    \centering
    \begin{tabular}{@{} c @{ } c @{ } c @{ } c @{ }}
      \includegraphics[width=.19\linewidth]{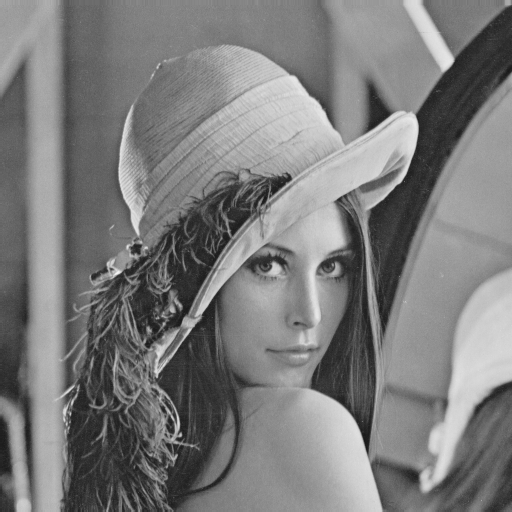}&
      \includegraphics[width=.19\linewidth]{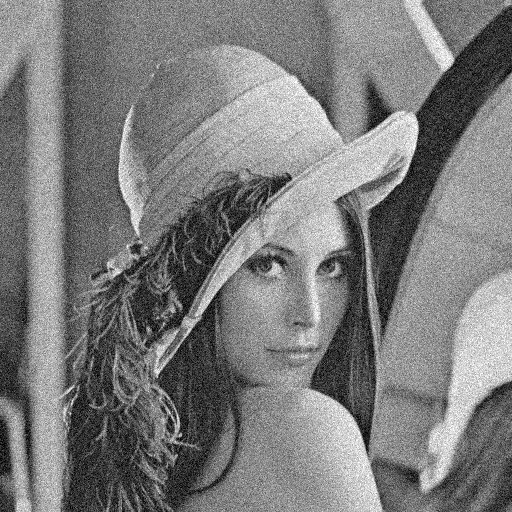}&
      \includegraphics[width=.19\linewidth]{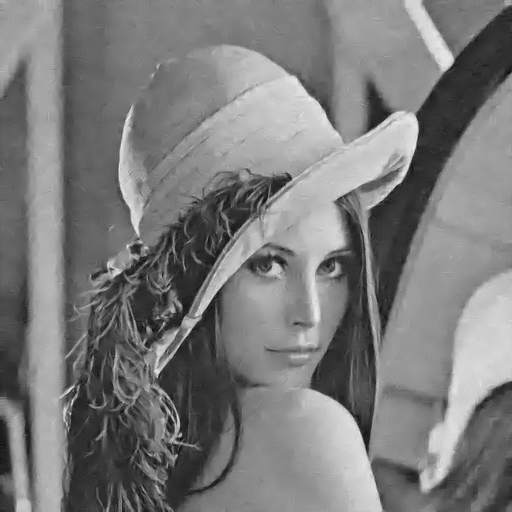}&
      \includegraphics[width=.19\linewidth]{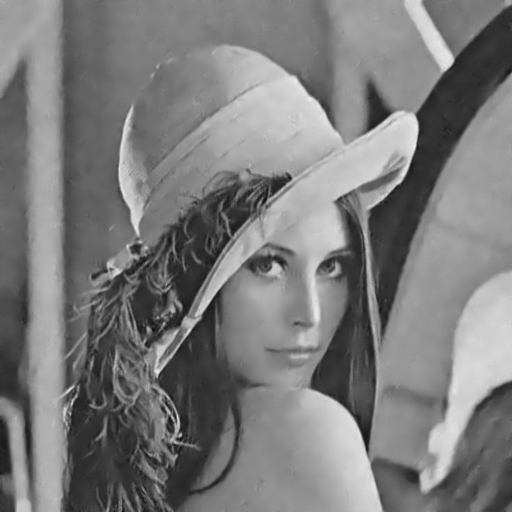}\\
      \end{tabular}
        \caption{Denoising of {\em Lena}. From left to right: original, noisy, our method denoising,KSVD denoising.}
      \label{fig:denoise}
\end{figure*}

    As can be seen in Fig.~\ref{fig:denoise} and Table~\ref{table1}, the learned CD generalizes well and the reconstruction is competitive both qualitatively and quantitatively to the results of KSVD denoising.

\begin{figure}[h]
\hspace{0.5in}
    \includegraphics[width=50mm,scale=0.8]{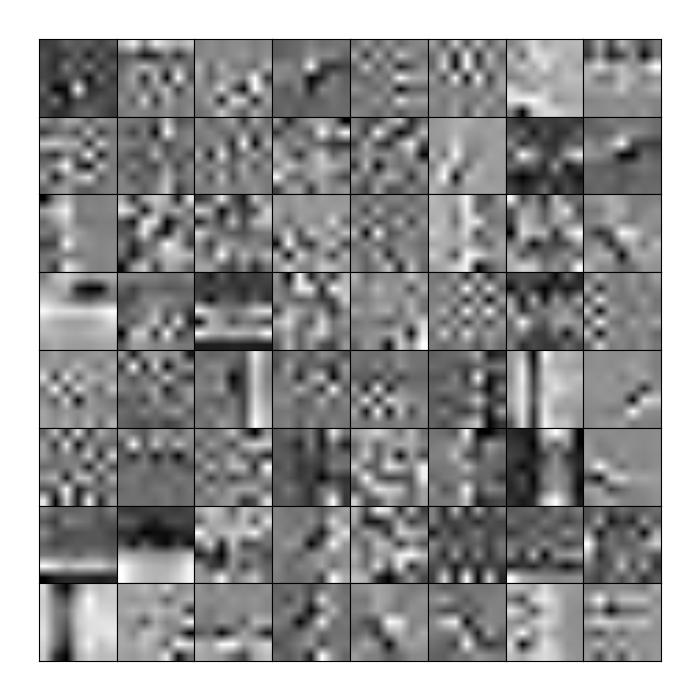}
    \caption{Convolutional dictionary learned on denoising task containing $64$ $7\times7$ filters with gabor and high-pass structure. }
    \label{fig:convdict}
\end{figure}

We show the atoms of the learned CD in  Fig.~\ref{fig:convdict}. The learned CD does not have any image specific atoms but rather a mixture of high pass DCT like and Gabor like filters.
Table \ref{runtime} presents the average run time per image. We used KSVD's publicly available code \cite{rubinstein2008efficient}.
Our model is faster by an order of magnitude on a CPU and by two orders on a GPU.
\begin{table}[h]
\begin{tabular}{lccc}
 &ours CPU&ours GPU&KSVD CPU\\ \hline
runtime [sec]&0.81&0.03&4.21\\ \hline
\end{tabular}
\caption{Run time on the image denoising task.}
\label{runtime}
\end{table}

	\begin{table}
	   \centering
        \small
			\begin{tabular}{cccc}\hline
            Image&Proposed&KSVD\\\hline
            Lena   &$\mathbf{32.11}$&$32.09$\\
            House  &$32.55$&$\mathbf{32.7}$\\
            Pepper &$30.65$&$\mathbf{30.89}$\\
            Couple &$\mathbf{30.14}$&$30.05$\\
            Fgpr   &$27.44$&$\mathbf{28.48}$\\
            Boat   &$30.3$&$\mathbf{30.37}$\\
            Hill   &$\mathbf{30.23}$&$29.5$\\
            Man    &$\mathbf{30.29}$&$29.67$\\
            Barbara&$28.91$&$\mathbf{30.57}$\\
		    \end{tabular}
			\caption{Denoising PSNR results for $\sigma_n=20$}
	\label{table1}
	\end{table}

 	\subsection{Image inpainting}
    We further test our model on the inpainting problem, in which $\mathbf{y}=\mathbf{M} \odot \mathbf{x}$, where  $\odot$ is as an element wise multiplication operator and $\mathbf{M}$ is a binary mask such that $\mathbf{y}$ contains only part of the pixels in $\mathbf{x}$.
    Rewriting (\ref{eq:8}) while taking $\mathbf{M}$ into consideration, we have
    \begin{equation}\label{eq:13}
        \mathbf{y}=\mathbf{M} \odot \mathbf{D}*\mathbf{z},
    \end{equation}
    Thus, (\ref{eq:5}) becomes,
	\begin{equation}\label{eq:14}
	 		\mathbf{z}_{k+1} = S_{\lambda/L}(\mathbf{z}_k + \frac{1}{L}\mathbf{d} \star (\mathbf{M} \odot (\mathbf{x}-\mathbf{d}*\mathbf{z}_k))),
	 \end{equation}
     The ACSC version of (\ref{eq:14}) is given by
	  \begin{equation}\label{eq:15}
\mathbf{z}_{k+1} = S_{\theta}(\mathbf{z}_k + \mathbf{w_e} * (\mathbf{M} \odot (\mathbf{x}-\mathbf{w_d}*\mathbf{z}_k))).
	 \end{equation}
	 In our experiment $M(i,j) \sim Ber(0.5)$, thus, we randomly sample half of
	 the input pixels. The objective is to reconstruct $\mathbf{x}$. We took a pre-trained ACSC network on denoising task, plugged in it $\mathbf{M}$ to be the form of (\ref{eq:15}), and optimized it for the inpainting task over the PASCAL VOC \cite{pascal-voc-2012} dataset.
	 We compare our inpainting results to \cite{heide2015fast} and \cite{papyan2017convolutional} over the same test images used \cite{heide2015fast}. The image numbering convention is consistent to  \cite{heide2015fast}.
All test images are  preprocessed with local contrast as in \cite{heide2015fast}. Table.~\ref{table2} shows that our model produces competitive results to the ones of \cite{heide2015fast} and \cite{papyan2017convolutional}.
The main advantage of the proposed approach as can be seen in Table~\ref{runtime1} is its significant speed-up in running time (by more than three orders of magnitude).

\begin{table}[t]
\begin{tabular}{lcccc}
 &ours CPU&ours GPU&\cite{heide2015fast} CPU&\cite{papyan2017convolutional} CPU\\ \hline
 runtime [sec]& 0.6& 0.023& 163& 65.49\\ \hline

\end{tabular}
\caption{Run time on image inpainting.}
\label{runtime1}
\end{table}

\begin{table}
\centering
 \scalebox{0.8}{%
\small
\begin{tabular}{cccc}
\hline
    \small{Image}& \small{Heide et al.} & \small{Papyan et al.} & \small{Proposed} \\ \hline
    \small{1}&  \small{28.76}&\small{$\mathbf{29.55}$}&\small{28.84} \\
    \small{2}&  \small{31.54}&\small{31.63}&\small{$\mathbf{31.95}$}\\
    \small{3}&  \small{30.59}& \small{$\mathbf{30.76}$}&\small{30.48}\\
    \small{4}&  \small{27.41}&\small{26.8}&\small{$\mathbf{27.64}$}\\
    \small{5}&  \small{33.65}&\small{33.59}&\small{$\mathbf{33.81}$}\\
    \small{6}&  \small{33.03}&\small{27.7}&\small{$\mathbf{33.27}$}\\
    \small{7}&  \small{28.69}&\small{$\mathbf{30.61}$}&\small{28.33}\\
    \small{8}&  \small{30.39}&\small{$\mathbf{32.36}$}&\small{30.23}\\
    \small{9}&  \small{28.07}&\small{$\mathbf{28.68}$}&\small{28.10}\\
    \small{10}& \small{31.59}&\small{$\mathbf{33.46}$}&\small{31.65}\\
    \small{11}& \small{29.77}&\small{$\mathbf{33.5}$}&\small{29.60}\\

    \small{12}&  \small{26.40}&\small{26.45}&\small{$\mathbf{26.49}$} \\
    \small{13}&  \small{29.01}&\small{$\mathbf{30.95}$}&\small{28.86}\\
    \small{14}&  \small{29.48}& \small{$\mathbf{30.76}$}&\small{29.5}\\
    \small{15}&  \small{28.95}&\small{$\mathbf{29.44}$}&\small{28.28}\\
    \small{16}&  \small{29.59}&\small{$\mathbf{30.45}$}&\small{29.48}\\
    \small{17}&  \small{27.69}&\small{28.27}&\small{$\mathbf{28.65}$}\\
    \small{18}&  \small{$\mathbf{31.69}$}&\small{31.61}&\small{31.36}\\
    \small{19}&  \small{26.79}&\small{$\mathbf{28.57}$}&\small{26.88}\\
    \small{20}&  \small{31.93}&\small{32.01}&\small{$\mathbf{32.04}$}\\
    \small{21}& \small{28.81}&\small{$\mathbf{28.91}$}&\small{$\mathbf{28.91}$}\\
    \small{22}& \small{26.27}&\small{$\mathbf{27.36}$}&\small{26.47}\\

\end{tabular} %
}
\caption{Inpainting PSNR for \cite{heide2015fast}, \cite{papyan2017convolutional} and  ACSC on test-set from \cite{heide2015fast}.}
    \label{table2}
\end{table}
\footnote{source code can be found \url{https://github.com/benMen87/AppriximateConvolutionalSparseCoding}}

    \section{CONCLUSION}
    Approximate convolutional sparse coding as proposed in this work is a powerful tool.
    It combines both the computational capabilities and approximation power of convolutional neural networks and  the strong theory of sparse coding. We demonstrated the efficiency of this strategy both in the tasks of image denoising and inpainting.
    \vfill \pagebreak
    \bibliographystyle{IEEEbib}
    \bibliography{refrences}

\end{document}